\documentclass{ws-p10x7}
\usepackage[]{latexsym}
\usepackage[]{amsmath}
\newcommand{\into}{\ensuremath{\rightarrow}}

\begin{document}

\title{Concluding Remarks}

\author{Nicola Cabibbo}

\address{Dipartimento di Fisica, Università di Roma - La Sapienza\\
 INFN, Sezione di Roma
 \\E-mail: nicola.cabibbo@roma1.infn.it}

\twocolumn[\maketitle\abstract{
Important new results have been presented at this conference. The 
direct  violation of $CP$ in $K^{0}\rightarrow \pi+\pi$ has been 
firmly established in two independent experiments, NA48 at CERN and 
KTeV at Fermilab. Both Babar at SLAC and Belle at KeK have determined 
the  $CP$ violation in $B^{0}_{d}-\bar B^{0}_{d}$ oscillations through the 
study of the golden $K_{S}+\Psi$ decay mode. The observed   $CP$ 
violation agrees with the expectations of the Standard model, based
on the quark-mixing phenomenon. The first results of the Sudbury 
Neutrino Observatory, SNO, suggest that the long-lasting solar 
neutrino puzzle has been finally solved in terms of neutrino oscillations.
Results appeared after the conference which modify the theoretical 
prediction of the muon anomaly. This new result, if confirmed, 
would drastically reduce the significance of the discrepancy between the 
theoretically expected value for the  muon anomaly and the recent 
ressults of the Brookhaven experiment.
}]

\section{Introduction}\label{sec:intro}%1

One of the pleasures of this conference has been the chance to meet
Alberto Sirlin after many years. He reminded me of a recipe for the perfect
closing lecture I offered  him many (30+) years ago: ``You have to mention 
everybody who gave a talk!''

I am glad to have the recipe back after such a long time, but I will 
not be able to follow it. This meeting is 
rich with important results, among which two new examples of
$CP$ violation, the possible solution of the solar neutrino puzzle in 
terms of neutrino oscillations and a possible discrepancy between the 
recent measurement of the muon anomaly and theoretical predictions. I 
will concentrate my attention on these subjects.

The discussion of these very hot arguments should not however
make us forget many other excellent results presented at the 
conference; the field is indeed progressing on a very wide front.

Among the many experimental results presented at the conference I was 
particularly impressed by those obtained at Hera by the ZEUS and 
H1 collaborations, which graphically demonstrate the unification of 
weak and electromagnetic interactions: at low  $Q^{2}$ the cross section 
for charged current events, $e\rightarrow \nu$,  is many 
orders of magnitude smaller than that for neutral current events,
$e\rightarrow e$, dominated by e.m. interactions; at high 
$Q^{2}$ the two cross sections inch closer and become proportional, as 
predicted by the standard model.

The Hera groups have presented a detailed determination of the scaling 
violation in deep inelastic scattering, allowing an extensive check on 
the predictions obtained from perturbative QCD, and an accurate 
determination of $\alpha_{s}(M_{Z})$. With the advent of more
accurate (NNLO) calculations the new experimental results will allow a 1\% 
precision in this important parameter.

\section{$CP$ Violation}\label{sec:CPVio}%2

\subsection{Quark mixing and $CP$ Violation}\label{subsec:mix}%2.0

The charged-current weak interactions of hadrons are described by the 
unitary matrix\cite{cab,km} $\mathbf{V}$ 
$\; (\mathbf{V^{\dagger}}\mathbf{V}=1)$. 
With only two families, e.g. in a world without beauty (or $t$ quarks),
$\mathbf{V}$ can always be reduced to a real form, so that $CP$ is 
necessarily conserved.

With three families the matrix $\mathbf{V}$ can be expressed in terms of four 
parameters:

\begin{eqnarray}
    &
     \mathbf{V} = 
     \begin{vmatrix}
              V_{ud} &V_{us}&V_{ub}\\
              V_{cd} &V_{cs}&V_{cb}\\
              V_{td} &V_{ts}&V_{tb}\\
    \end{vmatrix}\approx\nonumber\\
   &
    \begin{vmatrix}
              1-\lambda^{2}/2 &\lambda&A\lambda^{3}(\rho-i\eta)\\
              -\lambda &1-\lambda^{2}/2&A\lambda^{2}\\
              A\lambda^{3}(1-\rho-i\eta)&-A\lambda^{2}&1\\
    \end{vmatrix}
    \label{eq:Wolf}
\end{eqnarray}
where I have used the Wolfenstein~\cite{wolf} parametrization. 
A non-vanishing value for $\eta$ leads 
to the violation of $CP$ symmetry. With three quarks $CP$ 
conservation, $\eta=0$, is an exceptional case; $CP$ violation is the norm, 
obtained for any non-zero value of $\eta$.

Two of the 
parameters, $\lambda$ and $A$ are known with good precisions: 
$\lambda=\sin \theta$, where $\theta$ is my original mixing angle, 
is determined by $K_{l3}$ decays,
\begin{equation}
     \lambda=\sin\theta=0.2237 \pm 0.0033
    \label{eq:lambda}
\end{equation}

\begin{figure*}
%\epsfxsize160pt
\epsfxsize30pc
\figurebox{}{}{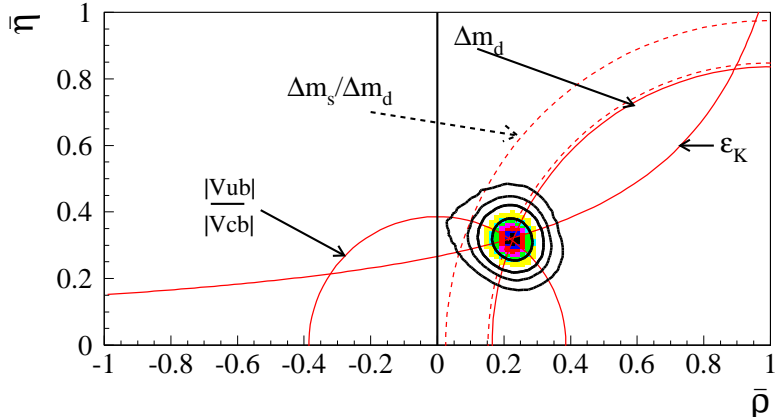}
\caption{Constraints on $\rho$, $\eta$ arising from  
$V_{ub}$, $\epsilon$ --- the $CP$ violating parameter in $K^{0}-\bar 
K^{0}$ mixing, and the $B^{0}_{d}-\bar B^{0}_{d}$ 
mixing parameter $\Delta m_{d}$.}
\label{fig:ciufit}
\end{figure*}

\begin{figure}%1
\epsfxsize120pt
\figurebox{}{}{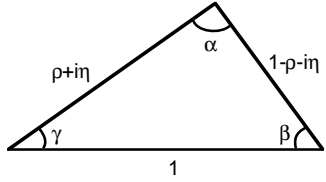}
\caption{The Unitarity triangle in the $\rho$ -- $\eta$ plane.}
\label{fig:utriangle}
\end{figure}

The rates of the allowed $B$ decays lead a determination of the $A$ 
parameter. From the analysis by Ciuchini et al. \cite{ciu} we have:
\begin{equation}
    A\lambda^{2}=V_{cb}= (41.0 \pm 1.6)\times 10^{-3} \, .
    \label{eq:Al2}
\end{equation}

The problem of determining the two remaining parameters, $\rho$ and $\eta$, 
is best seen in the light of the unitarity relation
    \begin{equation}
        V_{ud}V_{ub}^{*}+V_{cd}V_{cb}^{*}+V_{td}V_{tb}^{*}=0
        \label{eq:unit}
    \end{equation}
which can be represented as a triangular relation in the complex 
plane. The unitarity relation is automatically satisfied in the parametrization 
of Eq.~(\ref{eq:Wolf}), where it reduces, up to terms
$O(\lambda^{2})$,  to the triangle of figure~\ref{fig:utriangle}.
Since the area of the unitarity 
triangle is $\eta/2$, a non-flat triangle implies $CP$ violation.

The form of the unitarity triangle can
 be determined by measurements of $CP$ conserving quantities. 
The oscillation of $B^{0}_{d}$ mesons are dominated by 
graphs with virtual top quarks, so that the mass difference $\Delta 
m_{d}$ is proportional to $|1-\rho-i\eta|^{2}$, the length squared of 
one of the upper sides of the triangle. The length of the other 
side, $|\rho+i\eta|$, can be extracted from a determination of 
$V_{ub}$, e.g from a determination of the rates of the forbidden 
$b\rightarrow u$ leptonic transitions. These determinations point to a 
non-flat triangle, i.e. to the presence of a certain amount of $CP$ violation.
As a first check the values of $\rho,\eta$ so obtained agree well 
with the observed value of the $CP$ violating $\epsilon$ parameter
in $K^{0}-\bar K^{0}$ mixing. These different constraints on $\rho$ and 
$\eta$ are displayed in figure \ref{fig:ciufit}, and lead to the 
following estimates for $\rho$ and $\eta$:

\begin{equation}
\rho=0.224 \pm 0.038  ,~\eta=0.317 \pm 0.040  
\label{eq:eta1}
\end{equation}
What is perhaps more relevant is the fitted value 
for $\sin(2\,\beta)$,

\begin{equation}
\sin(2\,\beta)= 0.698 \pm 0.066\, ,\quad 
\label{eq:eta2}
\end{equation}
since this parameter is directly accessible through a study of $CP$ 
violation in the ``golden decay mode''~\cite{BigiSanda} of $B^{0}$ mesons, 

\begin{equation}
(B^{0}_{d}\; \text{or}\; \bar B^{0}_{d})\rightarrow K_{S}+\Psi
\label{eq:gold}
\end{equation}

In his presentation to this conference, C. 
Sachrajda has emphasized the central role of Lattice QCD simulations 
in the determinations of the CKM parameters. Lattice QCD was used for 
evaluating the $B$ parameter for $K$ mesons, needed in the 
prediction of $\epsilon$ in terms of $\rho$ and $\eta$, and again 
for determining the decay and mixing parameters $f_{B}$ and $B_{B}$ 
for both the $B_{d}$  and $B_{s}$
mesons, parameters which are needed for the determination of the two 
mass differences $\Delta m_{d}$ and  $\Delta m_{s}$. The 
present simulations are executed within the quenched approximation,
due to the limited  computer power available today, while more 
accurate simulations will be possible with the advent of 
teraflop class computers.

\subsection{New results on $CP$ violation in $B^{0}$ decays} \label{sec:four}
Results on $CP$ violation in the ``golden mode'' of Eq. \eqref{eq:gold}
have been presented at this conference both by the Babar experiment at 
SLAC and the Belle experiment at KeK. The results are in reasonable 
agreement among themselves and lead to a value of $\sin(2\,\beta)$ 
which is in good agreement  with the prediction of Eq. 
\eqref{eq:eta2}, a remarkable confirmation of the hypothesis that 
$CP$ violation phenomena arise from complex elements of the 
$\mathbf{V}$ matrix.
The ``golden'' character of $B^{0}_{d}\rightarrow K_{S}+\Psi$ derives from 
the fact that the final state is a $CP$ eigenstate, and that this decay mode 
is dominated by a $CP$ conserving tree
diagram. Any $CP$ violation observed in this mode must, to 
an excellent approximation, be attributed to $B^{0}_{d}-\bar B^{0}_{d}$ mixing. 

The interpretation of  $CP$ violation in $B^{0}_{d}-\bar B^{0}_{d}$ mixing is 
uniquely simple, since this mixing is dominated by a single diagram  
whose phase is easily seen to be $\exp(2\,i\,\beta)$. The measurement of 
$CP$ violation effects in the decays of Eq. \eqref{eq:gold} can be 
directly interpreted as a measurement of the $\beta$ angle in the 
unitarity triangle of figure \ref{fig:utriangle}.
 
The situation in $B^{0}_{d}-\bar B^{0}_{d}$ mixing is very different 
from that in $K^{0}-\bar K^{0}$ mixing, which is dominated by a $CP$ 
conserving diagram, $CP$ violation arising from a second smaller 
diagram. Contrary to $B^{0}_{d}$ case, obtaining information on the 
mixing matrix from the measurement of $CP$ 
violation in $K^{0}$ mixing (the $\epsilon$ parameter) 
requires a complex theoretical analysis and one must,
as noted above, recur to lattice QCD simulations to obtain an estimate 
of one of the required parameters.

The values of $\sin(2\,\beta)$ presented here by the two experimental 
groups\cite{Babar,Belle} are:

\begin{eqnarray}
\sin(2\beta)= 0.59 \pm 0.14_{\text{stat}} \pm 0.05_{\text{syst}}
     & (\text{Babar})\nonumber\\
\sin(2\beta)= 0.99 \pm 0.14_{\text{stat}} \pm 0.06_{\text{syst}}
     & (\text{Belle})\nonumber\\
     &
\label{eq:etaexp}
\end{eqnarray}
These are impressive results: each of them by itself establishes the 
existence of $CP$ violation in $B^{0}$ decays to many $\sigma$'s. It 
is remarkable that two rather different experiments at different 
accelerators and in different Laboratories were able to obtain  
results of comparable accuracy within a few days of each other.

Three previous measurements of $\sin(2\beta)$, obtained by CDF at 
Fermilab\cite{cdfsin2b}, and by Aleph and Opal at 
CERN\cite{lepsin2b}, have 
larger errors but are generally compatible with the new results, which 
however supersede earlier preliminary results by the same groups.
Combining the five extant results Ahmed Ali\footnote{
I am grateful to dr. Ali for providing this result. He commented: 
``the chi square of the fit is 5.2 to be compared with the expected 
$\chi=4$. The chi-square is not great but acceptable.''
} obtains the ``world average''

\begin{equation}
\sin(2\,\beta)= 0.79 \pm 0.12\, ,\quad 
\label{eq:etaworld}
\end{equation}
in excellent agreement with the theoretical prediction in 
Eq. \eqref{eq:eta2}.

\subsection{Direct $CP$ Violation in $K^{0}\rightarrow \pi\pi$;
          $\epsilon'/\epsilon$}

Two experimental groups, NA48 \cite{NA48}  
at CERN and KTeV \cite{KTeV} at Fermilab,
have presented new determinations of the direct 
violation of  $CP$ in the decays $K^{0}\rightarrow \pi^{+}\pi^{-}$, 
$K^{0}\rightarrow \pi^{0}\pi^{0}$, through a measurement of 
$Re(\epsilon'/\epsilon)$: 

\begin{eqnarray}
Re(\epsilon'/\epsilon)= (15.3 \pm 2.6)10^{-4} & NA48 \nonumber\\
Re(\epsilon'/\epsilon)= (20.7 \pm 2.8)10^{-4} & KTeV 
\label{eq:epsprime}
\end{eqnarray}
The two new results are in good 
agreement, and each of them is many $\sigma$'s away from 
$Re(\epsilon'/\epsilon)= 0$, so that the presence of direct 
$CP$ violations in $K^{0}$ decays is firmly established. 
The new results are in rough agreement with the previous result by
NA31 at CERN and with that obtained by he E731 experiment at Fermilab
(which was however compatible with $\epsilon'=0$).

The new world average is 
\begin{equation}
Re(\epsilon'/\epsilon)= (17.2 \pm 1.8)10^{-4} 
\label{eq:epsworld}
\end{equation}
This is in general agreement with the theoretical evaluations, which 
are however not excessively precise. They are normally quoted as 
``from a few $10^{-4}$ to $\approx 2\times 10^{-3}$''. The problem is 
that the direct violation of $CP$ in $K^{0}\rightarrow \pi\pi$ decays
involves ``penguin diagrams'' which are at present very hard to evaluate 
in Lattice QCD. Progress is expected in this direction with the 
advent on the one side of new high-performance parallel computers and 
on the other of new algorithms for the simulation of  low-mass quarks.

\section{Neutrino oscillation: the Solar neutrino puzzle solved?}

The solar neutrino puzzle has been with us for over thirty years, 
since the Davis chlorine experiment~\cite{Davis} detected only about a 
third of the neutrinos expected on the basis of the current solar model
\cite{Bahcall68}.

The deficit of solar neutrinos has over the years been confirmed 
by the Kamiokande and Super-Kamiokande water detectors, and by the 
gallium experiments, GALLEX, SAGE and GNO. At the same time the solar 
model has been refined, and tightened with the help of data on 
heliosismography, so that we can exclude that the neutrino deficit can 
find its explanation in some modification of the solar model itself.
We can refer the reader to the recent review~\cite{bahcPins} by
Bahcall, Pinsonneault and Basu. 

Already in 1968 Bruno 
Pontecorvo~\cite{pontecorvo68} proposed that 
a deficit in solar neutrinos could signal the presence of neutrino 
oscillations. An important theoretical development was the 
realization that the coherent interaction with the solar matter
can modify the neutrino oscillations~\cite{wolfSun},
and that this can give rise to resonant transformation between 
neutrino species even for small mixing angles~\cite{MiSm}. 
The coeherent interaction of solar neutrinos with the 
bulk matter of earth could give rise to \emph{day-night} effects which 
could be explored by real-time detectors such as Kamland or Borexino.

Many questions remained open: are oscillations real? are the 
oscillations confined to the known neutrino flavours, or do they 
involve also new flavours (called sterile neutrinos)
which do not partake of neutral current 
interactions, and do not therefore contribute to the $Z^{0}$ width?

Results obtained at the Sudbury Neutrino Observatory (SNO)\cite{SNO},
presented 
at this conference, seem to give a positive answer to this two 
questions: The solar neutrinos which arrive at the Earth behave
as a mixture of $\nu_{e}$ and other \emph{active} neutrinos, i.e. 
$\nu_{\mu}$ and $\nu_{\tau}$.

The principle of the SNO experiment is to compare the 
rate of charged current inverse beta decay events, which can only 
arise from $\nu_{e}$'s, and of neutrino-electron scattering events, to 
which also $\nu_{\mu}$'s and $\nu_{\tau}$'s can contribute. The ratio 
of the two type of events,
\begin{equation}
    \frac{CC}{ES}\propto\frac{\nu_{e}}{\nu_{e}+0.14(\nu_{\mu}+\nu_{\tau})}
    \label{eq:SNOtrick}
\end{equation}
can be used to deduce the total number of neutrinos, which can than be 
compared with the solar model prediction. The SNO collaboration has 
obtained an accurate determination of the $CC$ rate, while their 
determination of the $ES$ rate is not accurate enough to establish the
existence, in the solar flux at the Earth, of a fraction of $\nu_{\mu}$ 
and $\nu_{\tau}$ . They can however use the determination of 
the $ES$ rate obtained \cite{SuperKa} at Super-Kamiokande which has the 
required precision. The two measurements are mainly sensitive to neutrinos 
in the same energy band, the $B^{8}$ neutrinos, so that their combination is 
meaningful. The neutrino fluxes determined through $CC$ and $ES$ events 
differ by more than three standard deviations, thus giving a strong 
support for the existence of neutrino oscillations:
\begin{equation}
    \Phi^{ES}_{S-K}-\Phi^{CC}_{SNO}= (0.57 \pm 0.17)10^{6}cm^{-2}s^{-1}
    \label{eq:SNOResult}
\end{equation}
The two measurements, together with 
Eq. \eqref{eq:SNOtrick}, determine the total flux of $B^{8}$ 
neutrinos,
\begin{equation}
    \Phi= (5.44 \pm 0.99)10^{6}  \text{cm}^{-2}\text{s}^{-1},
    \label{eq:SNOB8}
\end{equation}
which is in excellent agreement with the solar model predictions.
The solar neutrino gap seems to have closed.

In the next few years we expect important results from both 
SuperKamiokande and SNO. Two new experiments will 
give important contributions to the unravelling of the solar neutrino
problem:
\begin{itemize}
    \item  KamLand will study oscillations in reactor neutrinos with
    a sensitivity sufficient to confirm or exclude the --- now 
    favoured --- LM solution for neutrino oscillations.

    \item  Borexino will be able to observe in real time the flux of 
    the low-energy $Be^{7}$ solar neutrinos. This will allow
    refined studies of
    day/night and seasonal effects.
\end{itemize}

The new SNO data favour \cite{Fogli,BahcallSNO} large mixing angle 
oscillation solutions, which opens the way to the possibility of $CP$ and $T$ 
violation in  neutrino oscillations:
\begin{center}
\begin{eqnarray*}
  CP: (\nu_1\into \nu_2)&\leftrightarrow(\bar\nu_1\into\bar \nu_2)\\
  T\;\;: (\nu_1\into \nu_2)&\leftrightarrow(\nu_2\into \nu_1)
\end{eqnarray*}
\end{center}
Disappearence experiments $(\nu_{1}\into \nu_{1})$ 
cannot display $CP$ or $T$ violations

Exploring $CP$ or $T$ violations requires superbeams, or better a dedicated 
neutrino factory, which is also the first step for a muon collider. 
Both possibilities have been discussed during this conference.

I cannot resist quoting from a paper I wrote in 1978 \cite{CabibboNuPC}:
\begin{center}
  ``maximal neutrino mixing \underline{requires} $CP$ violation''  
\end{center}
By maximal I mean that all matrix elements of the lepton 
mixing matrix $\mathbf{V_{L}}$ should have equal size.

Since $\mathbf{V_{L}}$ is unitary, the requirement of maximal mixing 
has essentially a unique solution which is necessarily complex:

\begin{equation*}
    \mathbf{V_{L}}=
         \begin{vmatrix}
              1 &x&x^{2}\\
              x &x^{2}&1\\
              x^{2}&1&x\\
    \end{vmatrix};\quad
   x=\exp(\frac{2\pi i}{3})
\end{equation*}

We are probably far from this solution, but perhaps not \underline{very} far.

\section{The muon anomaly: signal for new physics?}
James Miller presented here the recent results of the 
Brookhaven measurement~\cite{g-2recent} of the muon magnetic anomaly. 
The new world average,
\begin{equation}
a_{\mu}^{Exp}= (1165920.3 \pm 1.5)\times 10^{-9} 
\label{eq:worldanomaly}
\end{equation}
disagrees with the theoretical prediction~\cite{Czarnecki:1998nd} by 
nearly three standard deviations.  
\begin{eqnarray}
&a_{\mu}^{Th}= (1165915.96 \pm .67)\times 10^{-9} \label{eq:thanomaly} \\
& a_{\mu}^{Exp} - a_{\mu}^{Th} = (4.3\pm1.6)\times 10^{-9}
\label{eq:ThExp}
\end{eqnarray}
The Brookhaven collaboration expects to be able to decrease the 
experimental error by nearly a factor three in the near future.
Already in its present state the discrepancy seems serious and has 
stimulated a multitude of theoretical papers which examine  different 
possible implications of this discrepancy, which would clearly be a 
signal for new physics.

The interesting aspect is that the discrepancy is relatively large; 
by comparison the contribution to the muon anomaly of electroweak 
effects --- diagrams with virtual $Z^{0}, W^{\pm}$ bosons contribute 
a correction
\begin{equation}
\delta^{EW} a_{\mu}= (1.51 \pm 0.04)\times 10^{-9} 
\label{eq:EWanomaly}
\end{equation}
In order to explain a discrepancy of $(4.3\pm1.6)\times 10^{-9}$ one 
would need new physics at relatively low energies, in other words this 
discrepancy looks as excellent news for the forthcoming LHC experiments.
It is also clear that in view of the importance of a possible 
discrepancy both the experimental analysis and the theoretical 
computations must be submitted to to the most careful scrutiny. 

The theoretical prediction of the muon anomaly is the sum of diagrams 
with virtual leptons, photons and intermediate vector bosons,
\begin{eqnarray}
&\text{QED}\quad 116584706(3)&\times 10^{-11}
   \label{eq:muonQED}\\
&\text{EW} \quad \quad \quad \quad \quad \; 151(4) &\times 10^{-11}
\end{eqnarray}
and  diagrams which include  virtual hadrons, further divided in 
$\alpha^{2}$ and $\alpha^{3}$ diagrams which include hadron corrections 
to the photon propagator (PP), and diagrams with hadronic light by 
light (LL) subdiagrams. These diagrams are the main sources 
of the theoretical error,
\begin{eqnarray}
&\text{PP; $\alpha^{2}$} \quad \quad 6924(62)&\times 10^{-11}\\
&\text{PP; $\alpha^{3}$} \quad\quad -100(6)&\times 10^{-11}\\
&\text{LL; $\alpha^{3}$} \quad\quad -85(25)&\times 10^{-11}
   \label{eq:muonLL}
\end{eqnarray}
The numbers reported in eqs. (\ref{eq:muonQED} -- \ref{eq:muonLL}) are those 
used by the Brookhaven collaboratin in their analysis.

The hadron corrections to the photon propagator can  be related 
to the total cross section for hadron production in electron positron 
collisions~\cite{CabibboGatto}; their contribution to the muon anomaly
can then be expressed 
as an integral, with a suitable kernel, over the cross section for 
$e^{+}\,e^{-}\rightarrow$ hadrons. 
The most important part of this 
contribution and of its  error arises from the low energy ($\leq 1 
GeV$) region. As an alternative to low energy $e^{+}\,e^{-}$ data one 
can use, via the CVC relation,  data on the $\tau$ decay into hadrons,
which are at present more accurate~\cite{Davier}.
A number of evaluations of the hadronic photon propagator 
contribution to the muon anomaly
have appeared in recent times, with slightly different 
results~\cite{moreHadrons} and slightly different evaluations of the 
error. 

Since the hadron correction to the photon propagator
is safely anchored to experimental 
data on $e^{+}e^{-}$ collisions and $\tau$ decays, the error on this 
contribution to the muon anomaly will improve in the next few years. 
Particularly promising is 
the advent of the KLOE experiment at the DA$\Phi$NE $\phi$ factory, which 
should provide a new standard of accuracy for  $e^{+}e^{-}$ cross 
sections in the low energy region.

The situation of the light by light  (LL) hadronic contribution is very 
different, as we have not found in this case a way to evade the complexities of 
hadron physics by relating this contribution to other measurable  
phenomena. Waiting for a frontal attack on this contribution 
using lattice QCD (not easy) we must be satisfied with models whose 
accuracy is difficult to estimate.

The current evaluations of the hadronic light by light contributions 
to the muon anomaly \cite{kinoshita,bijnens} are based on models of 
the light pseudoscalar mesons and their interactions at low energy - 
chiral perturbation theory or the extended Nambu - Jona Lasinio model.
The dominant hadronic LL contribution turns out to be the one mediated by 
a single intermediate neutral pion. 

After the LP01 conference was concluded, 
a new calculation of the $\pi^{0}$ contribution~\cite{knecht,deRafael}
to the muon anomaly reached a very suprising conclusion: while previous 
calculations had found a negative sign,  the new 
result, recently confirmed by an independent 
computation~\cite{Blokland}, found a positive  sign. 
The authors of the two complete evaluations of 
the  LL contributions  \cite{kinoshita,bijnens}  have been able to
identify the origin of what they now see as a sign error in the
previous computations and  have presented new evaluations for the 
overall LL contributions, which are respectively ~\cite{kinoshita2,bijnens2}
$(89\pm 16)\times 10^{-11}$
and $(83\pm 32)\times 10^{-11}$. The effect of this 
sign change is a reduction of the discrepancy to less than 2 standard 
deviations. 

Quite apart from the question of sign of the $\pi^{0}$ contribution, 
which only arose after the conference, we must note a detailed 
criticism~\cite{melnikovAnomaly} by K. Melnikov, who argues that 
the contribution of quark-loop light by light diagrams has been 
underestimated in ref. \cite{kinoshita,bijnens} In  referring the interested reader
to Melnikov's paper for the details of his argument, I note that 
this would further reduce the discrepancy between theoretical 
evaluations and the experimental result. Although the authors of 
ref. \cite{bijnens} do not agree with this argument,
it is clear that, in order to  calculate the muon anomaly with a precision
comparable with that expected from the  Brookhaven experiment, 
the hadronic light by light contributions must be carefully re-evaluated.
The forthcoming measurements of hadron production in low energy 
electron-positron collisions should lead to an improved evaluation of 
the contribution of  hadronic corrections to the photon propagator. 

\section{We are not alone!}
While all this has being going on, cosmologists\ldots

We heard in the talks by Halzen and Turner of the 
exciting progress our neighbours are making in Cosmology and
Astrophysics. With the recent results on the
cosmic background anisotropy cosmologists are confirming their own 
Standard Model. 

The new results on the cosmic background arise from a serendipitous 
use of the antartic winds which circulate around the South Pole: a 
balloon released from an antartic station comes back close to the same spot 
in obout a month. A balloon  circling the South Pole  can be very 
competitive with a satellite: the launch is by far 
less expensive, and the payload does not need to meet the high 
standards and associated cost in both money and time that a space 
launch requires. 

In the Boomerang flights the cosmic background has been studied with 
unprecedented resolution. The angular resolution of the recent data 
corresponds to spherical harmonics of $\approx 1000$. In this range three peaks 
are evident in the power spectrum, which fit very well the 
expectation for a Big-Bang universe  at $\Omega=1$, i.e. a flat 
universe, whose energy density is much larger than the ``observed'' 
baryon density which would correspond to  $\Omega\approx 0.05$.
This offers futher evidence for the conclusion that most of the 
matter in the universe is ``dark matter'', most probably ``cold dark 
matter'', i.e. matter constituted of relatively heavy particles, 
which have decoupled from ``normal'' matter early in the history of 
the universe. 

While it is clearly the task of astronomers and cosmologist to 
ascertain the geometry and history of the universe, the task of 
attempting the detection of these slow particles coasting along in the 
present universe falls to the high energy community. Many experiments 
are now underway for the detection of the weak interacting cold 
matter, and might bear fruit in the coming years.

\section{Conclusions and acknowledgements}

This conference has been enlivened by many exciting results. 
Will the next one be even better? It is a tall order, but beautiful things 
are brewing. 

After decades in 
which $CP$ violation was established in a single process, the 
$K^{0}-\bar K^{0}$ oscillations, we  now 
have two more well established examples, the first in $B^{0}-\bar B^{0}$ oscillations 
and the second a direct violation in $K^{0}\rightarrow \pi\,\pi$. The first is 
important for the light it sheds on quark mixing and the Standard 
Model in general, while the measurement of $\epsilon'/\epsilon $ has 
a very special impact because, apart from it being in general 
agreement with the still imprecise expectations of the Standard Model, 
it definitely excludes the ``Superweak'' models of $CP$ violation.

We can look forward to new results on $CP$ violation: Babar and Belle 
should be able to establish new examples of direct violation in $B$ 
decays and the KLOE experiment at DA$\Phi$NE should offer a 
determination of  $\epsilon'/\epsilon $ which is logically 
independent from those presented by the NA48 and KTeV experiments.

The other very exciting development comes fro the SNO results which 
corroborate the conclusion that the solar neutrino puzzle will find 
its solution in neutrino oscillations. Here we expect important 
results in the near future from both SNO and Super-Kamiokande, but also 
from experiments which are now approaching the data-taking phase, 
in the first instance Kamland and Borexino.

The new results by SNO reinforce the 
proposal that neutrino oscillations are characterized by large mixing 
angles, and this opens up a very exciting possibility of detecting  
$CP$ violation effects in neutrino oscillations. The detection of 
these effects will however require new neutrino beam facilities, 
which have been discussed during LP01. A first attempt could be 
carried out with superbeams while more detailed studies will require 
the availability of full---flegded neutrino factories.

The success of this conference is certainly the merit of the many 
research groups who have contributed important new results, and of the 
many physicists who have contributed well prepared and well documented 
presentations of the new data, but LP01 would not have succeeded 
without the efforts of the organizers and of the many young 
people who have devoted so much time and efforts  to its success.

I am particularly grateful to Juliet Lee Franzini and Paolo Franzini, 
who invited me to give these concluding remarks to a very exciting 
conference whch has turned out to be a real turning point in the kind 
of physics I have been working on for many years.

To everybody who participated in the conference I would like to 
present my best wishes that we may all be working very hard, and be 
ready to surprise each other when we meet  in 2003 for the next 
Lepton Photon Conference.

\end{document}